\documentclass[aps,reprint,amsmath,amssymb,nobibnotes,superscriptaddress]{revtex4-1}

\usepackage{graphicx}
\usepackage{dcolumn}
\usepackage{bm}

\begin{document}

\preprint{APS/123-QED}


\title{C$_4$N$_3$H monolayer: A novel two-dimensional organic Dirac material with high Fermi velocity}


\author{Hongzhe Pan}
\affiliation{National Laboratory of Solid State Microstructures, Collaborative Innovation Center of Advanced Microstructures and Jiangsu Provincial Key Laboratory for Nanotechnology, Nanjing University, Nanjing 210093, China}
\affiliation{School of Physics and Electronic Engineering, Linyi University, Linyi 276005, China}
\author{Hongyu Zhang}
\affiliation{Department of Physics, East China University of Science and Technology, Shanghai 200237, China}
\author{Yuanyuan Sun}
\affiliation{National Laboratory of Solid State Microstructures, Collaborative Innovation Center of Advanced Microstructures and Jiangsu Provincial Key Laboratory for Nanotechnology, Nanjing University, Nanjing 210093, China}
\affiliation{School of Physics and Electronic Engineering, Linyi University, Linyi 276005, China}
\author{Jianfu Li}
\affiliation{School of Physics and Electronic Engineering, Linyi University, Linyi 276005, China}
\author{Youwei Du}
\affiliation{National Laboratory of Solid State Microstructures, Collaborative Innovation Center of Advanced Microstructures and Jiangsu Provincial Key Laboratory for Nanotechnology, Nanjing University, Nanjing 210093, China}
\author{Nujiang Tang}
\email[]{Corresponding author: tangnujiang@nju.edu.cn}
\affiliation{National Laboratory of Solid State Microstructures, Collaborative Innovation Center of Advanced Microstructures and Jiangsu Provincial Key Laboratory for Nanotechnology, Nanjing University, Nanjing 210093, China}


\date{\today}

\begin{abstract}
Searching for two-dimensional (2D) organic Dirac materials, which have more adaptable practical applications in comparing with inorganic ones, is of great significance and has been ongoing. However, only two kinds of these materials with low Fermi velocity have been discovered so far. Herein, we report the design of an organic monolayer with C$_4$N$_3$H stoichiometry which possesses fascinating structure and good stability in its free-standing state. More importantly, we demonstrate that this monolayer is a semimetal with anisotropic Dirac cones and very high Fermi velocity. This Fermi velocity is roughly one order of magnitude larger than that in 2D organic Dirac materials ever reported, and is comparable to that in graphene. The Dirac states in this monolayer arise from the extended $\pi$-electron conjugation system formed by the overlapping 2\emph{p}$_z$ orbitals of carbon and nitrogen atoms. Our finding opens a door for searching more 2D organic Dirac materials with high Fermi velocity.
\end{abstract}

\pacs{}

\maketitle

\section{\label{sec:level1}introduction}

Following with the great development of graphene \cite{1} and topological insulators \cite{2}, an emerging field of ¡°Dirac physics¡± is being established for investigating the quantum relativistic properties of a class of special materials with Dirac cones, namely Dirac materials \cite{3}. Such materials exhibit linear electronic band dispersion at the Fermi level, that is Dirac band, and thus have charge carriers behaved like massless Dirac fermions. The unique Dirac bands endow these materials with many novel phenomena in electronic transport, such as ballistic charge transport and high carrier mobility \cite{4}, Klein tunneling \cite{5}, various quantum Hall effects \cite{1,6,7}, \emph{etc}. These specific transport properties offer Dirac materials a wide range of promising applications in high-speed low-dissipation devices. In addition, the continuous reduction in the size of devices highly needs to develop low-dimensional materials. Thus, two-dimensional (2D) Dirac materials are much more desirable for the applications in nanoscale integrated circuits.

Theoretically, based on the generalized von Neumann-Wigner theorem \cite{8}, at least three conditions are required to achieve Dirac bands in 2D materials: (i) specific symmetries, (ii) proper parameters and (iii) appropriate Fermi level and band overlap \cite{9}. The hexagonal symmetry was widely believed to the most favorable for the existence of Dirac bands due to the fact that many 2D Dirac materials are observed in hexagonal lattice. However, it is not a necessary precondition for the presence of 2D Dirac materials. For example, 6,6,12-graphyne \cite{10}, square-MoS$_2$ \cite{11} and \emph{Pmmn}-boron \cite{12} have rectangular lattices while also possess Dirac cones. Besides symmetries, proper structure parameters such as bond lengths and angles, which are related to the crystal lattice, are also required. Furthermore, the Fermi level should precisely lie at the Dirac points, while there should be not any other bands overlap at the Fermi level. This condition is required to ensure the experimental observation and practical applications of the novel properties particular to the Dirac cones. Unfortunately, 2D Dirac materials are found to be very rare due to these rigorous conditions \cite{9}. For instance, among hundreds of 2D materials, only graphene \cite{1,4}, phagraphene \cite{13}, graphynes \cite{10}, silicene and germanen \cite{14}, borophenes \cite{12}, FeB$_2$ monolaye \cite{15}, \emph{etc}., have been confirmed to be 2D Dirac materials. Moreover, most of the existing 2D Dirac materials are inorganic compounds. In general, organic materials have the additional advantages of mechanical flexibility and tunable properties in comparison with inorganic ones. Therefore, it is of great practical significance to search for more 2D organic Dirac materials with the development of ¡°Dirac physics¡±.

Actually, an extensive search for 2D organic Dirac materials has been ongoing. However, to the best of our knowledge, only two kinds of 2D organic systems have been discovered to be Dirac materials so far. The first successfully realized example is the 2D layered organic conductor $\alpha$-(BEDT-TTF)$_2$I$_3$ (BEDT-TTF = bis(ethylenedithio)-tetrathiafulvalene) which turns into a Dirac material with a pair of titled Dirac cones when applying a high hydrostatic pressure above 1.2 GPa \cite{16,17}. The other one is a theoretical design about some 2D conjugated polymers. These polymers can be theoretically constructed by replacing the insulating connectors (1,3,5-triazine, \emph{etc}.) of conventional semiconducting 2D covalent organic frameworks (COFs) with the conductive ones (trivalent carbon atoms, \emph{etc}.) \cite{18}. Unfortunately, both of the two kinds of 2D organic materials have very low Fermi velocities, which are roughly one order of magnitude lower than that in 2D inorganic Dirac materials \cite{4,10,12,13,14,15}. It is known that a high Fermi velocity is favorable to the transport property of a Dirac material \cite{19}. Thus, designing stable 2D organic Dirac materials with high Fermi velocities is urgent and of great practical significance.

Inspired by the experimental synthesis of several 2D carbon nitride sheets, such as graphitic carbon nitride (g-C$_3$N$_4$) \cite{20,21}, C$_2$N holey 2D crystal (C$_2$N-\emph{h}2D) \cite{22} and C$_3$N sheets \cite{23} based on condensation reactions, we design a novel 2D organic material named as C$_4$N$_3$H monolayer by its stoichiometry. This organic monolayer has a intriguing structure with evenly distributed heart-shaped angstrom-scale pores and rather good dynamic, thermal and mechanical stabilities in its freestanding state. More interestingly, we demonstrate that this organic monolayer is a 2D Dirac material with anisotropic Dirac cones and a very high Fermi velocity of $1.1\times10^{6}$ m s$^{-1}$. Remarkably, this Fermi velocity is roughly one order of magnitude larger than that in 2D organic Dirac materials ever reported \cite{17,18}, and is even comparable to that in 2D inorganic Dirac materials \cite{4,10,12,13,14,15}. The Dirac points in this monolayer locate at off-symmetry points between the $\Gamma$ and K points, and arise predominantly from the overlapping 2\emph{p}$_z$ orbitals of carbon (C) and nitrogen (N) atoms. In addition, we also comment on the experimental feasibility of producing the predicted C$_4$N$_3$H monolayer and propose two hypothetical synthesis routes.

\section{\label{sec:level1}Computational Details}

Structural optimization, energy, density of states (DOS), band structure, electron localization function (ELF) \cite{24} and deformation charge density calculations based on density functional theory were performed in the framework of the generalized gradient approximation with the PBE functional \cite{25} using Vienna ab initio simulation package (VASP) \cite{26,27}. Projector-augmented plane wave approach \cite{28} was used to represent the electron interaction. A 500 eV energy cutoff of the plane-wave basis sets and the first Brillouin zone sampled with a $45\times45\times1$ Monkhorst-Pack \emph{k}-points grid were adopted in the structural relaxations and self-consistent calculations. In all the calculations, a vacuum distance of 15 \AA\ was applied along the perpendicular direction to ensure negligible interaction between adjacent layers. All atoms were allowed to relax without any constraint in the geometric optimizations until the total energy change was less than $1.0\times10^{-5}$ eV and the force on each atom was smaller than 0.001 eV \AA$^{-1}$. According to the crystal symmetry, the band structure of the C$_4$N$_3$H monolayer was calculated along the special lines connecting the following high-symmetry points: $\Gamma$ (0, 0, 0), K (0.4, 0.4, 0), M (0.5, 0, 0), R (0.6, -0.4, 0), S (0.5, -0.5, 0), $\Gamma$ (0, 0, 0), and M (0.5, 0, 0) in the \emph{k}-space. Moreover, to confirm the existence of Dirac states, the band structure of the C$_4$N$_3$H monolayer was also recomputed respectively by the more accurate Heyd-Scuseria-Ernzerhof (HSE06) functional \cite{29} and the spin-orbit coupling (SOC) effect.

The finite displacement method, as implemented in the phonopy code \cite{30}, was employed to calculate the phonon dispersion curves of the C$_4$N$_3$H monolayer. During the phonon spectrum calculation, a $4\times4\times1$ supercell was employed and the force constant matrix was determined by the VASP. First-principles molecular dynamics (FPMD) simulations, as also performed in the VASP, were employed to evaluate the thermal stability of this monolayer. The initial configuration with a $4\times4\times1$ supercell was annealed at different temperatures of 300, 500 and 1000 K, respectively. The temperature was controlled by the Nos\'e-Hoover thermostat \cite{31}. At each temperature, FPMD simulations in NVT ensemble lasted for 30 ps with a time step of 2.0 fs. A kinetic energy cutoff of 450 eV and the PBE functional were employed in all FPMD simulations.

\section{\label{sec:level1}Results and Discussion}
\subsection{\label{sec:level2}Geometric structure of the C$_4$N$_3$H monolayer}

\begin{figure}
\includegraphics[width=8.5cm]{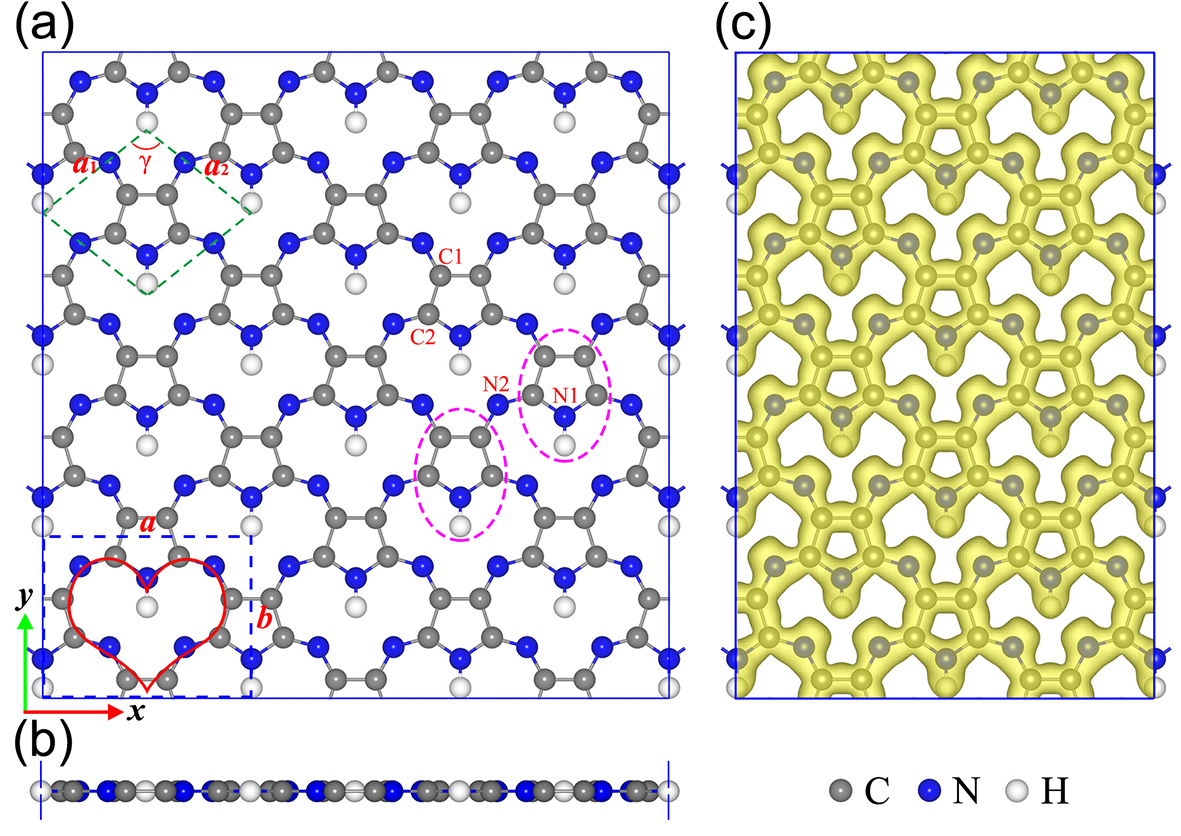}
\caption{\label{fig1} (a) Top and (b) side views of the optimized geometric structure of the C$_4$N$_3$H monolayer. The primitive cell and conventional cell are respectively denoted by the green and blue dashed lines; \textbf{\emph{a}$_1$} and \textbf{\emph{a}$_2$} represent the lattice vectors of the primitive cell and $\gamma$ is the angle between them; \emph{a} and \emph{b} are the lattice constants of the conventional cell. The red solid line is the connection line among the certain atoms around the angstrom-scale pore in the conventional cell. C1 and C2 denote different C atoms, and N1 and N2 represent different N atoms. (c) Isosurface of the ELF of C$_4$N$_3$H monolayer plotted with the value of 0.5.}
\end{figure}

Figures \ref{fig1}(a) and \ref{fig1}(b) respectively present the top and side views of the geometric structure for the optimized C$_4$N$_3$H monolayer. As shown in Fig. \ref{fig1}(a), this monolayer has a rhombic primitive cell (represented by the green dash lines) with the fully relaxed lattice constants \emph{a}$_1$ = \emph{a}$_2$ = 4.77 \AA, $\gamma = 104.5^{\circ}$ (angle between the \textbf{\emph{a}$_1$} and \textbf{\emph{a}$_2$} lattice vectors). Obviously, its primitive cell consists of four C atoms, three N atoms and one H atom. The correspondingly optimized lattice constants of the transformed rectangular conventional cell [denoted by the blue dash lines in Fig. \ref{fig1}(a)] with the same symmetry are \emph{a} = 7.54 \AA\ and \emph{b} = 5.84 \AA, respectively. This structure has the symmetry of $C_{2v}^{14}$ and belongs to the space group of \emph{Amm2}. It is obvious that angstrom-scale pores evenly distribute in the structure of the C$_4$N$_3$H monolayer. Intriguingly, if we link the atoms around the angstrom-scale pore in its conventional cell one by one, the connecting line shows a perfect heart-shape [denoted by the red solid line in Fig. \ref{fig1}(a)]. By anatomizing this intriguing structure, one can find that this monolayer is actually constructed by using N atoms to link the framework of pyrrole molecules [represented by the pink dash lines in Fig. \ref{fig1}(a)]. As shown in Fig. \ref{fig1}(a), we label the N atoms in the framework of pyrrole molecules as N1, and the ones sited at the linking positions as N2. There are also two kinds of C atoms, i) the C atoms with the nearest neighbors of two C atoms and one N atom (labeled as C1), and ii) the rest of C atoms which has one C atom and two N atoms as the nearest neighbors (labeled as C2). The C1--C1 and C1--C2 bond lengths (1.43 and 1.46 \AA, respectively) are only slightly larger than that in graphene (1.42 \AA, a typical bond length for \emph{sp}$^2$ C--C bonds), and are noticeably smaller than 1.54 \AA\ (the standard value of C--C bond length for \emph{sp}$^3$ hybridization) \cite{32}. The lengths of C1--N2, C2--N1, and C2--N2 bonds in this monolayer are respectively about 1.33, 1.38 and 1.31 \AA, very similar to that in the already-synthesized g-C$_3$N$_4$ (\emph{ca}. 1.33 and 1.41 \AA) \cite{21} and C$_2$N-\emph{h}2D (1.33 \AA) \cite{22} which have stable structures with \emph{sp}$^2$ hybridization bonds. More detailed information about the structural properties is summarized in Fig. 1 of the Supplemental Material \cite{33}. In addition, it is noteworthy that the C$_4$N$_3$H monolayer has an exactly planar structure, as shown in Fig. \ref{fig1}(b). This exactly planar structure and the features of bond lengths and bond angles imply that the chemical bonds in this monolayer ought to be covalent bonds with \emph{sp}$^2$ hybridization.

\begin{figure}
\includegraphics[width=6.8cm]{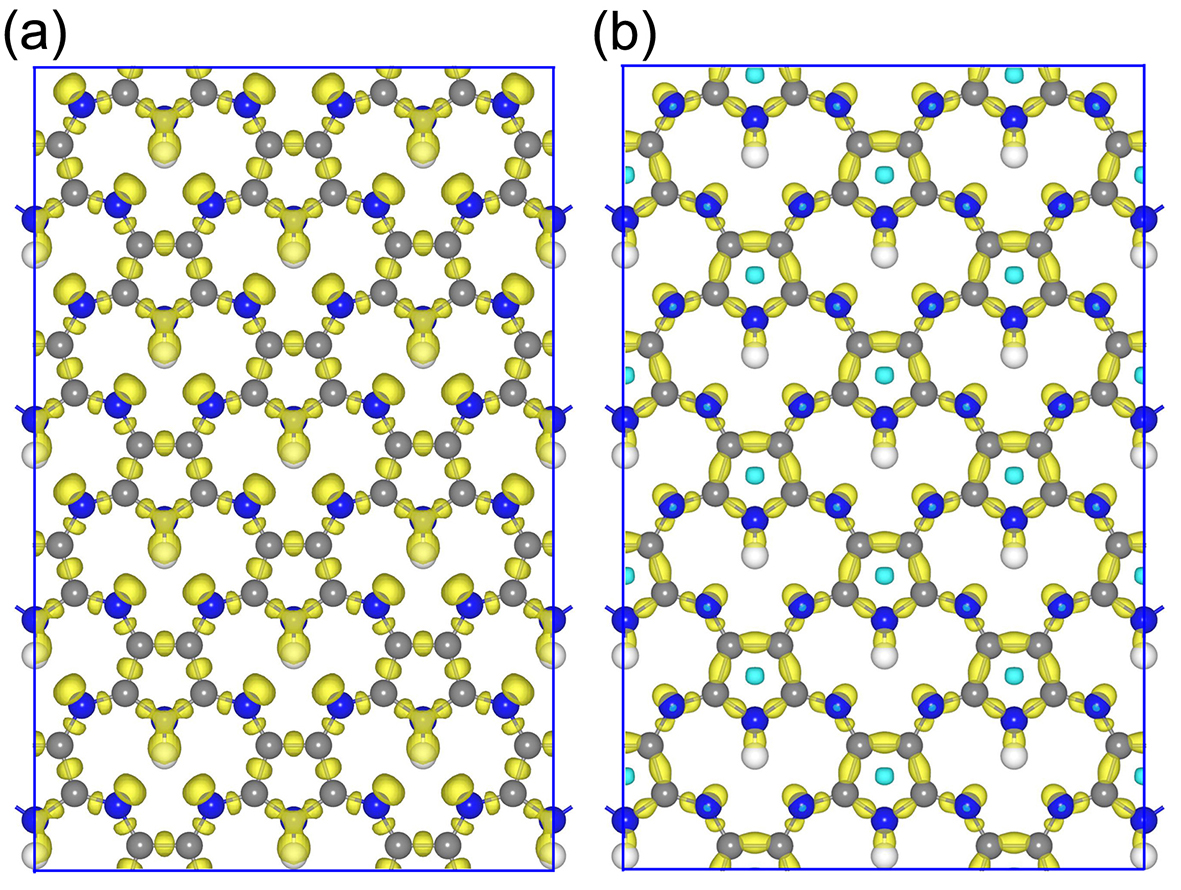}
\caption{\label{fig2} (a) ELF plotted with the value of 0.8. (b) Deformation charge density of the C$_4$N$_3$H monolayer. Yellow and cyan refer to electron accumulation and depletion regions, respectively. The isovalue of deformation charge density is 0.3 e \AA$^{-3}$.}
\end{figure}

To confirm this conjecture and further elucidate the bonding nature of the C$_4$N$_3$H monolayer, we then calculated the ELF to analyze its electron distributions. As known, ELF can be described in the form of isosurface in real space with isovalues ranging from 0 to 1. The region with 1 indicates the strong covalent electrons or lone-pair electrons, the region close to 0 implies the area with low electron density, and the region with an isovalue of 0.5 is an area with homogeneous electron gas. As shown in Fig. \ref{fig1}(c), the electron gas is well distributed and delocalized at the whole region of this monolayer network, which can electronically stabilize the 2D framework. To highlight the in-plane bonding states, we also plotted the isosurface of ELF for this monolayer with an isovalue of 0.8 in Fig. \ref{fig2}(a). It is found that the ELF localization centers clearly locate at the middle of the C--C, C--N and N--H bonds, indicating that the bonds have strong covalent electron states with $\sigma$-like \emph{sp}$^2$ hybridization. The $\sigma$ bonds between C, N and H atoms can also be evidenced by the deformation electronic density of the C$_4$N$_3$H monolayer [Fig. \ref{fig2}(b)]. The deformation electronic density is defined as the total electronic density of this monolayer excluding those of isolated atoms. Clearly, electrons are well localized over the C--C, C--N and N--H bonds, confirming the conclusion obtained from the ELF and the conjecture we proposed above. In addition, the remaining valence electrons of carbon and nitrogen atoms form the delocalized $\pi$ network, like the case of graphene. According to the Bader charge analysis \cite{34,35}, C1, C2, N1, N2 and H atoms in the C$_4$N$_3$H monolayer respectively possess $+0.55$, $+1.08$, $-1.28$, $-1.25$ and $+0.55$ $|e|$ charge. This level of charge transfer between C and N atoms is similar to that in stable C$_2$N-\emph{h}2D monolayer which has been successfully synthesized \cite{22,36}, suggesting that the C$_4$N$_3$H monolayer is likely to have the similar stability.

\subsection{\label{sec:level2}Stability of the C$_4$N$_3$H monolayer}

To evaluate the feasibility of experimental synthesis and the stability of the C$_4$N$_3$H monolayer, we first calculated its cohesive energy ($E_{\text{coh}}$) defined by $E_{\text{coh}}=(n_{\text{C}}E_{\text{C}}+n_{\text{N}}E_{\text{N}}+n_{\text{H}}E_{\text{H}}-E_{\text{C$_4$N$_3$H}})/(n_\text{C}+n_\text{N}+n_\text{H})$, where $E_\text{C}$, $E_\text{N}$, $E_\text{H}$ and $E_\text{C$_4$N$_3$H}$ are the calculated total energies of isolated C, N and H atoms, and C$_4$N$_3$H monolayer, respectively; $n_\text{C}$, $n_\text{N}$ and $n_\text{H}$ are the number of C, N and H atoms in the supercell of this monolayer, respectively. According to our computations, the cohesive energy is 7.88 eV per atom. This value is evidently larger than that of the already-synthesized borophene (5.87 eV per atom) \cite{37}, silicene (4.01 eV per atom) \cite{38} and phosphorene (4.67 eV per atom) \cite{39}, implying the high stability and synthetic feasibility of the C$_4$N$_3$H monolayer from the viewpoint of energy level.

\begin{figure}
\includegraphics[width=6cm]{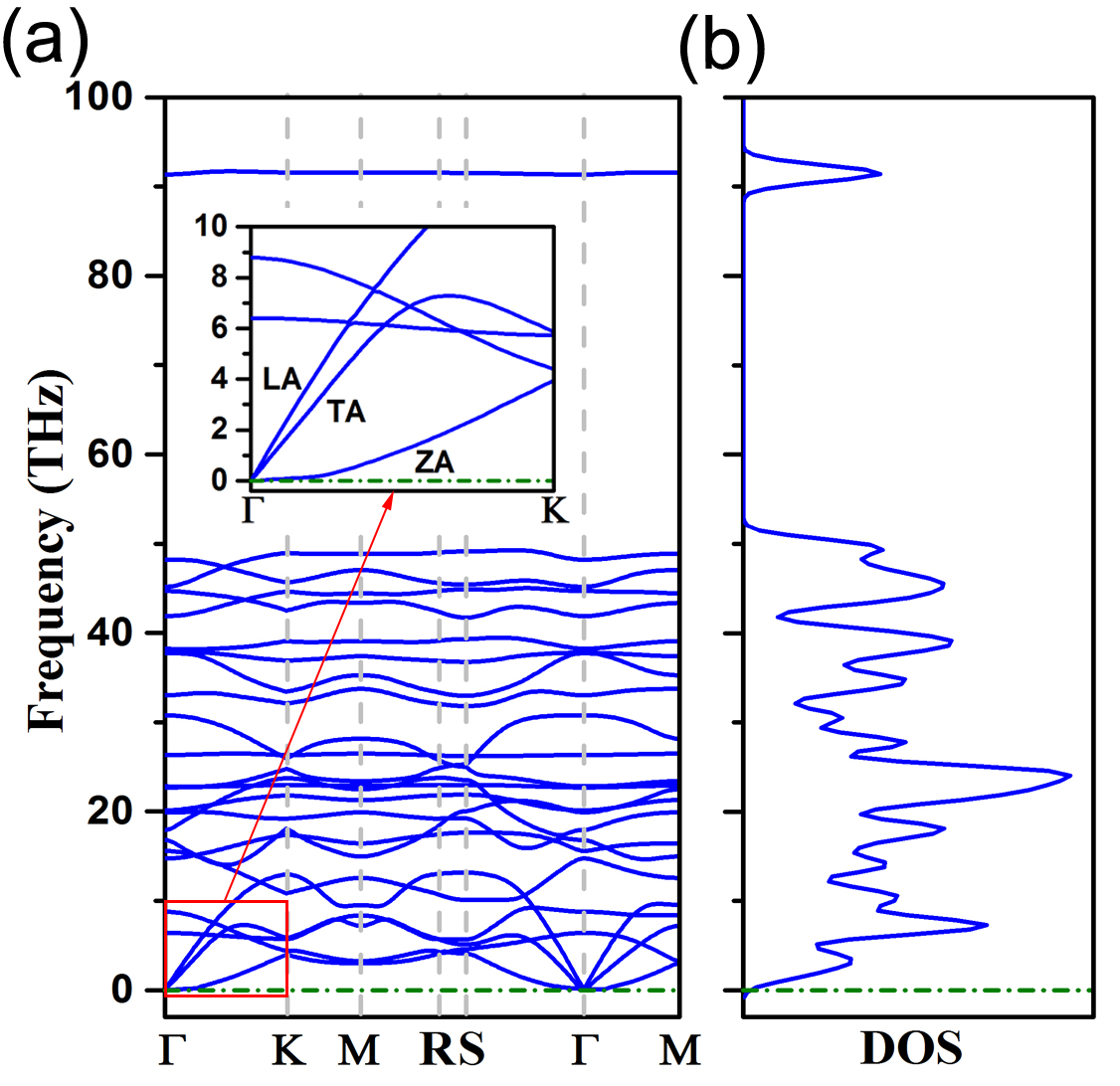}
\caption{\label{fig3} (a) Phonon dispersion curves and (b) total phonon DOS of the C$_4$N$_3$H monolayer. Inset is the enlarged drawing of red rectangle part in the phonon dispersion curves.}
\end{figure}

The stability of this monolayer can be further confirmed by its phonon dispersion curves and phonon DOS. As shown in Fig. \ref{fig3}(a), there is no sign of imaginary phonon mode in the phonon spectrum along the highly symmetric points in the entire Brillouin zone. In detail, there are eight atoms in the primitive cell of the C$_4$N$_3$H monolayer, thus its phonon spectrum has twenty-four phonon bands, including three acoustic branches and twenty-one optical branches. The three acoustic branches are respectively transverse acoustic (TA) and longitudinal acoustic (LA) branches corresponding to vibration within the plane, and the other one (ZA) corresponding to vibration out of plane [inset of Fig. \ref{fig3}(a)]. It can be seen that in contrast to the linear dispersion for the TA and LA branches, the frequency of the ZA branch shows a quadratic dispersion near the $\Gamma$ point. It is worth mentioning that this type of quadratic dispersion of ZA branch is a characteristic feature of the phonon dispersion curves in monolayered or layered crystals, e.g., graphene \cite{40}, graphite \cite{41} and other layered compounds \cite{15,42,43}. As shown in Fig. \ref{fig3}(b), the total phonon DOS also reveals that no phonon with imaginary frequency is found in this monolayer, which agrees very well with its phonon spectrum. These results demonstrate the good dynamic stability of the C$_4$N$_3$H monolayer.

\begin{figure}
\includegraphics[width=8.5cm]{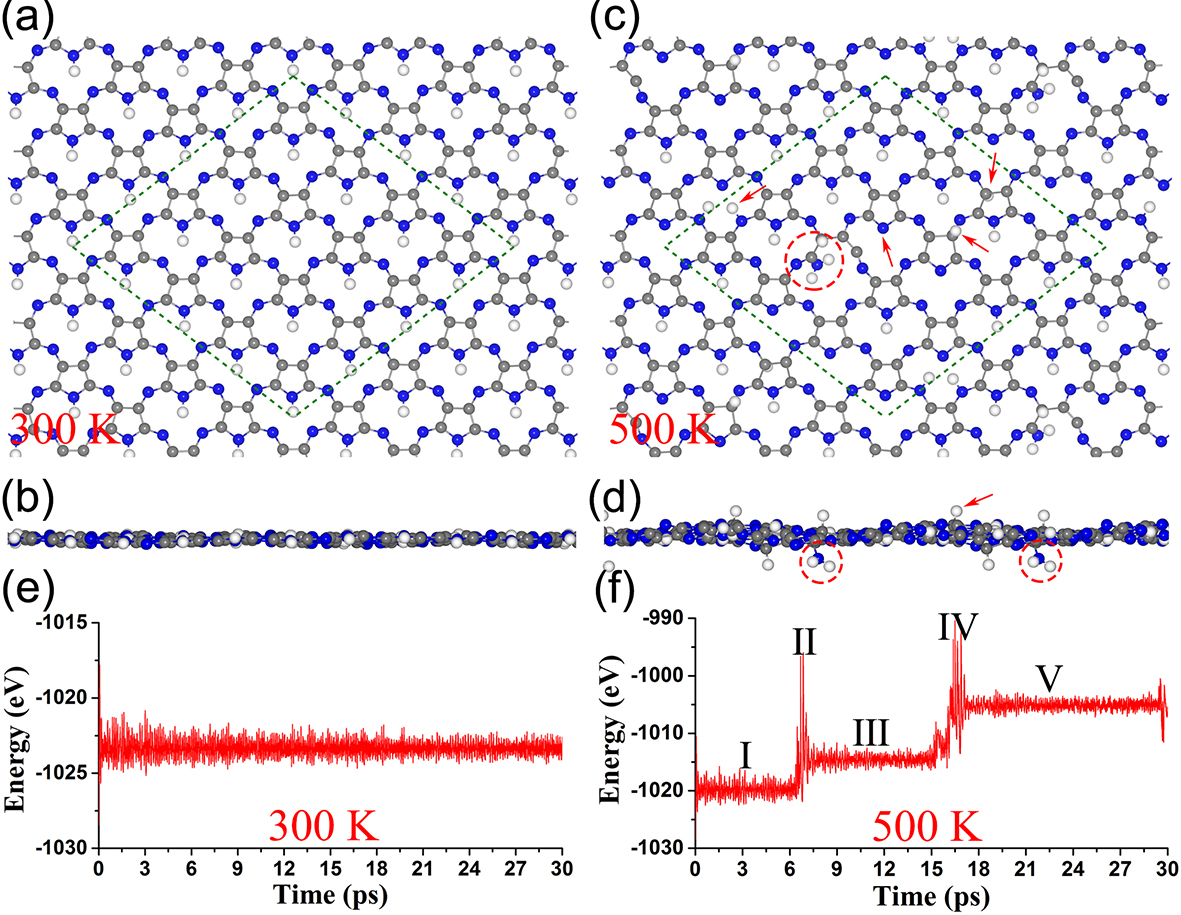}
\caption{\label{fig4} (a) Top and (b) side views of snapshots for the equilibrium structures of the C$_4$N$_3$H monolayer at the end of 30 ps FPMD simulations under the temperature of 300 K; (c) and (d) are similar respectively to Figs. 4(a) and 4(b), but for another FPMD simulation at the temperature of 500 K. The green dashed lines denote the $4\times4\times1$ supercell used in the FPMD simulations. The red arrows and circles display the migration of hydrogen atoms and the break of C--N bonds, respectively. (e) and (f) are the fluctuations of total energies with respect to FPMD simulation times at 300 K and 500 K, respectively.}
\end{figure}

Moreover, to further evaluate the thermal stability of the C$_4$N$_3$H monolayer, we performed FPMD simulations using a $4\times4\times1$ supercell containing 64 C atoms, 48 N atoms and 16 H atoms. The initial configuration was annealed at different temperatures of 300, 500 and 1000 K with a time step of 2 fs. As shown in Figs. \ref{fig4}(a) and \ref{fig4}(b), the snapshots of the geometry structure at the end of 30 ps simulations clearly reveal that the C$_4$N$_3$H monolayer can maintain its structural integrity except for some thermal fluctuations at the temperature of 300 K. The total energy of the simulated system can reach equilibrium quickly at 300 K [see Fig. \ref{fig4}(e)], verifying the above result from the viewpoint of energy level. This result can be understood by the fact that the binding energies of the C--C, C--N and C--H bonds are larger than the thermal energy corresponding to room temperature, consistent with other 2D carbon nitride systems \cite{21,22,23,44}. Moreover, this slightly distorted structure can restore its initial planar configuration after complete atomic relaxation.

The structure of this monolayer after annealing at 500 K is shown in Figs. \ref{fig4}(c) and \ref{fig4}(d). Obviously, significant atomic rearrangement took place and the basal plane substantially disordered after thermal annealing [see the partial structures pointed by the red arrows and circles in Figs. \ref{fig4}(c) and \ref{fig4}(d)]. Further analysis of the FPMD simulations reveals the process of the structure collapse at this temperature , as shown in Fig. \ref{fig4}(f) and Fig. 2 of the Supplemental Material \cite{33}. Moreover, this monolayer immediately and completely collapsed at a higher temperature of 1000 K. Hence, combined with the above result that the C$_4$N$_3$H monolayer can maintain its structural integrity at room temperature, we can get the conclusion that this monolayer has a melting point between 300 and 500 K, revealing its decent thermal stability.

Considering the importance of the mechanical stability of a material for its applications, we also studied the mechanical properties of this organic monolayer by examining its elastic constants. It is known that there are four nonzero elastic constants for a 2D material, which are $C_{11}$, $C_{22}$, $C_{12}$ ($C_{21}$) and $C_{66}$, respectively. These elastic constants need to satisfy the criteria ($C_{11}C_{22}-C_{12}^2>0,C_{66}>0$) for a mechanically stable 2D sheet \cite{45,46}. The in-plane Young¡¯s modules along \textbf{\emph{a}} ($Y_a$) and \textbf{\emph{b}} ($Y_b$) directions can be expressed as $Y_a=(C_{11}C_{22}-C_{12}C_{21})/C_{22}$ and $Y_b=(C_{11}C_{22}-C_{12}C_{21})/C_{11}$. The elastic constants of the C$_4$N$_3$H monolayer computed are $C_{11}$ = 209.8 N m$^{-1}$, $C_{22}$ = 138.7 N m$^{-1}$, $C_{12}$ = $C_{21}$ = 71.5 N m$^{-1}$ and $C_{66}$ = 86.0 N m$^{-1}$. Clearly, these elastic constants satisfy the above criteria, denoting the good mechanical stability of this monolayer. Accordingly, $Y_a$ and $Y_b$ calculated respectively are 173.1 and 114.3 N m$^{-1}$, indicating that the C$_4$N$_3$H monolayer is mechanically anisotropic. The Young¡¯s modules of this monolayer are higher than those of already-synthesized phosphorene ($Y_a$ = 25.5 N m$^{-1}$ and $Y_b$ = 91.6 N m$^{-1}$) \cite{39,46}, suggesting its good mechanical property.

Considering that the C$_4$N$_3$H monolayer has comparable cohesive energy to other already-synthesized 2D materials, and dynamic, thermal and mechanical stabilities, we believe that it is viable for its experimental synthesis. It is well known that the condensation reaction is a valid method for producing exotic materials in polymer chemistry. Most importantly, recent developments in this field have already successfully synthesized a series of polymer materials, for instance, g-C$_3$N$_4$ \cite{20,21}, C$_2$N-\emph{h}2D \cite{22}, C$_3$N sheets \cite{23}, 2D COFs \cite{47,48}, metal-organic frameworks \cite{49}, \emph{etc}. These materials show the tremendous capability of modern chemistry to create novel 2D networks from custom-designed monomers. Inspired by this, we proposed two hypothetical routes to synthesize the C$_4$N$_3$H monolayer based on pyrrole molecules. Firstly, one should respectively nitride (or oxidize) pyrrole molecules to the certain products, and then, 2D C$_4$N$_3$H material can be synthesized \emph{via} the condensation reaction of these intermediate products (see Fig. 3 of the Supplemental Material \cite{33}). Herein, it should be noted that the real synthetic process must be much more complicated and difficult. However, it is worth expecting for the experimental realization of this novel material due to its unique electronic properties revealed in the next section.

\subsection{\label{sec:level2}Electronic structures of the C$_4$N$_3$H monolayer}

\begin{figure}
\includegraphics[width=8.5cm]{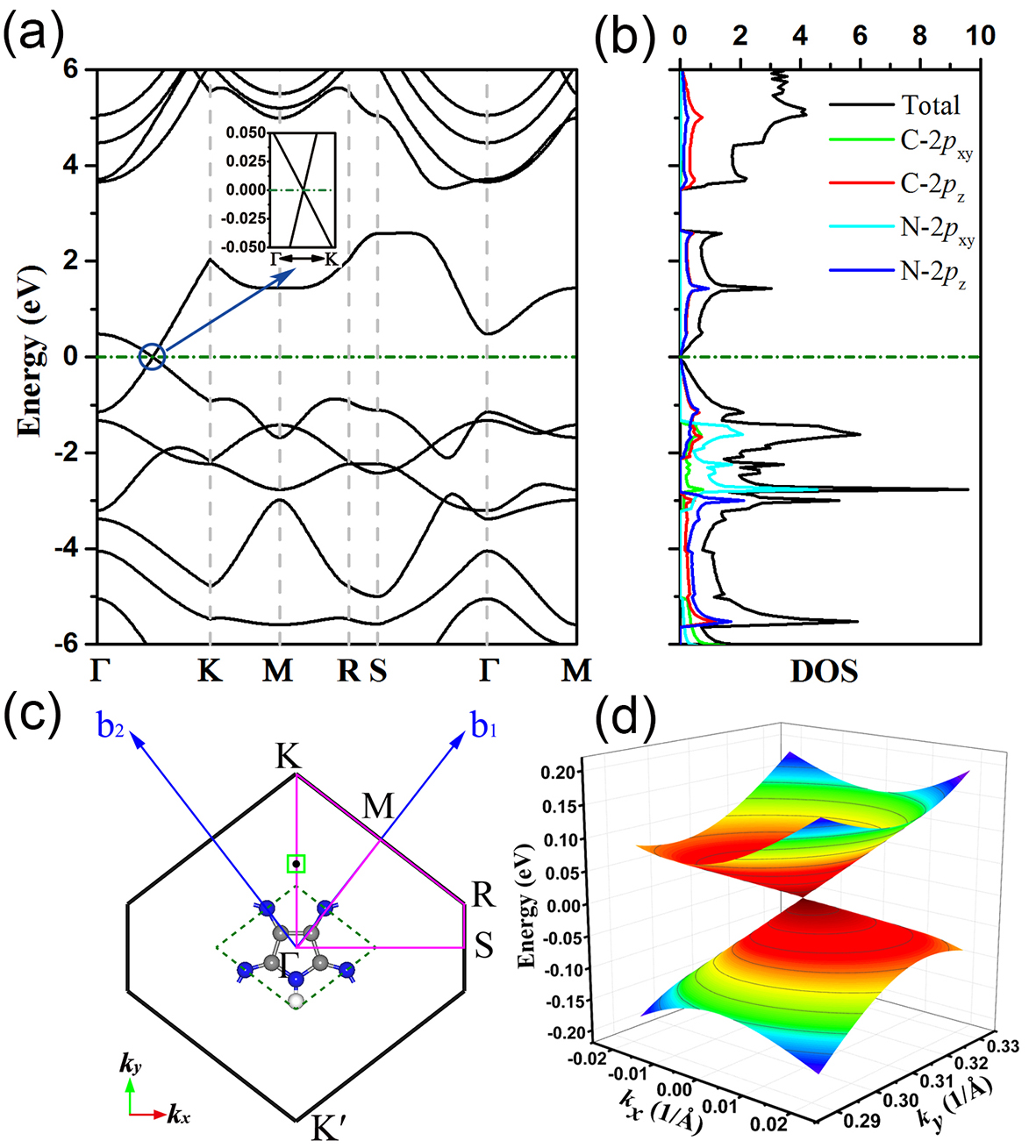}
\caption{\label{fig5} (a) Band structure and (b) TDOS and PDOS of the C$_4$N$_3$H monolayer. The Fermi level is assigned at 0 eV. Inset in Fig. \ref{fig5}(a) is the enlarged drawing of the bands in the vicinity of the Dirac point, which was calculated by using high-precision screening parameter of 0.0001 \AA$^{-1}$. (c) First Brillouin zone with the special \emph{k} points: $\Gamma$ (0, 0, 0), K (0.4, 0.4, 0), M (0.5, 0, 0), R (0.6, -0.4, 0), and S (0.5, -0.5, 0). \textbf{\emph{b$_1$}} and \textbf{\emph{b$_2$}} are reciprocal lattice vectors of the \emph{k}-space, while \textbf{\emph{k$_x$}} and \textbf{\emph{k$_y$}} are basis vectors of the rectangular coordinate system. The pink lines depict the high symmetric lines connected by the special k points while the black dot represents the position of the Dirac point. (d) Distorted Dirac cone formed by the valence and conduction bands in the vicinity of the Dirac point [the region displayed by the green square in Fig. \ref{fig5}(c)].}
\end{figure}

As a novel material with intriguing structure, its detailed electronic structure needs to be explored. Thus, we calculated the band structure, total DOS (TDOS) and projected DOS (PDOS) of the C$_4$N$_3$H monolayer. As shown in Fig. \ref{fig5}(a), one can see that this monolayer is a semimetal with the valence and conduction bands meeting in a single point at the Fermi level. The characteristics of linear bands and degenerate state at this point denote the appearance of Dirac states in the C$_4$N$_3$H monolayer. Different from the well-known graphene where Dirac points locate at the high-symmetry points (K and K$^{'}$ points), the Dirac point of the C$_4$N$_3$H monolayer locates on the path from the $\Gamma$ to K point. As a result, there are two symmetry-related Dirac points appearing in the entire first Brillouin zone [only one representative is shown in Figs. \ref{fig5}(a) and \ref{fig5}(c), as the two Dirac points are related by symmetry]. Consistently, its TDOS is zero at the Fermi level [Fig. \ref{fig5}(b)], supporting the presence of the Dirac point. Figure 5(c) shows the first Brillouin zone of the C$_4$N$_3$H monolayer with orthorhombic (\emph{Amm2}) structure, and further demonstrates that neither honeycomb structure nor hexagonal symmetry is prerequisite for the existence of Dirac cones.

To obtain deeper insight into the Dirac states of this organic monolayer, we further calculated the Dirac cone formed by the valence and conduction bands in the vicinity of the Dirac point. As shown in Fig. \ref{fig5}(d), its Dirac cone is distinctly distorted, similar to that of some inorganic Dirac materials such as phagraphene \cite{13}, 6, 6, 12-graphyne \cite{10} and \emph{Pmmn}-boronene \cite{12}. The linear dispersion curves of energy and momentum in both the \textbf{\emph{k$_x$}} and \textbf{\emph{k$_y$}} directions around the Dirac point suggest the zero effective mass of the carriers (electrons and holes) near the Fermi level. To examine the carrier mobility around the distorted Dirac cone, we then calculated the Fermi velocity ($v_F$) of the C$_4$N$_3$H monolayer using the formula of $v_F=(1/\hbar)\cdot(\partial E/\partial k)$, where $\partial E/\partial k$ is the slop of valence or conduction band near the Dirac point and $\hbar$ is the reduced Planck$^{'}$s constant. The slope of the bands in the \textbf{\emph{k$_x$}} direction is ¡À7.3 eV \AA, equivalent to a Fermi velocity $v_{Fx}=1.1\times10^6$ m s$^{-1}$, while in the \textbf{\emph{k$_y$}} direction, the slop of the bands equal to 6.2 eV \AA\ ($v_{Fy}=9.4\times10^5$ m s$^{-1}$) and -2.8 eV \AA\ ($v_{Fy}=4.3\times10^5$ m s$^{-1}$)(see Fig. 4 of the Supplemental Material \cite{33}). The largest Fermi velocity is comparable to that in graphene and other 2D inorganic Dirac materials \cite{10,12,13,14,15}, and is roughly one order of magnitude larger than that in 2D organic Dirac materials ever reported \cite{17,18}. It is noted that Fermi velocities in 2D organic polymers followed an approximately inverse exponential relation to their pore size, largely independent of the interconnecting oligomer \cite{18}. That is to say, the 2D organic Dirac polymer with the small pore size is favorable to the high Fermi velocity. Thus, the high Fermi velocity of our organic monolayer may attribute to its angstrom-scale pores. In addition, the anisotropy of the distorted Dirac cone with different slopes at the Dirac point in the \textbf{\emph{k$_x$}} and \textbf{\emph{k$_y$}} directions implies the direction-dependent electronic properties of the C$_4$N$_3$H monolayer, indicating its more flexible applications in contrast with graphene.

Additionally, the hybrid HSE06 functional \cite{29} and the SOC effect were respectively used to recalculate the band structure of the C$_4$N$_3$H monolayer for further confirming its Dirac states. As shown in Fig. 5(a) of the Supplemental Material \cite{33}, the dispersion of the valence and conduction bands at the Fermi level given by the HSE06 functional is very similar to that computed based on the PBE functional \cite{25} and no band gap can be identified, indicating that the above predicted intrinsic Dirac states in the C$_4$N$_3$H monolayer still survive to the hybrid HSE06 functional. On the other hand, we also demonstrate that the SOC effect on the electronic properties of this monolayer is negligible and the Dirac cone is still well preserved, which can be seen in Fig. 5(b) of the Supplemental Material \cite{33}. This is unsurprising due to the fact that C, N and H are all light elements, and the SOC effect on the nontrivial gap should be very small.

\begin{figure}
\includegraphics[width=8.5cm]{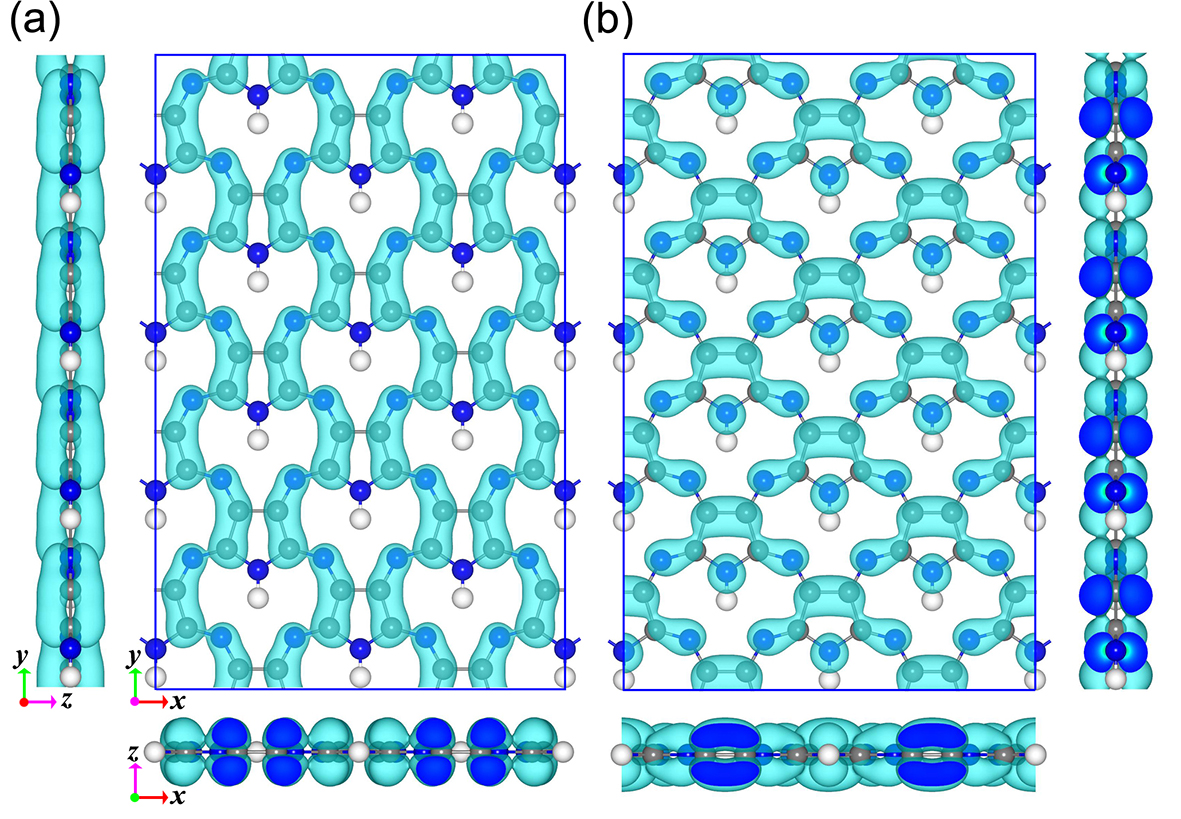}
\caption{\label{fig6}Top and side views of the isosurfaces of partial charge densities for (a) the highest valence band and (b) the lowest conduction band. The isovalue is 0.015 e \AA$^{-3}$.}
\end{figure}

Then, we explored the physical origin of the Dirac states in the C$_4$N$_3$H monolayer. Through analyzing the PDOS of this monolayer [Fig. \ref{fig5}(b)], one can see that both the occupied and unoccupied peaks near the Fermi level mostly originate from the 2\emph{p}$_z$ orbitals of C and N atoms. The calculated orbital-resolved band structures of this monolayer also show the same conclusion that the bands near the Fermi level are mainly contributed by 2\emph{p}$_z$ orbitals of the atoms (see Fig. 6 of  the Supplemental Material \cite{33}). Furthermore, we also calculated the band decomposed charge density at the Dirac point to visualize the above result. Clearly, the electron density distribution presents an obvious characteristic of 2\emph{p}$_z$ orbitals [see the side views of the isosurfaces plotted in Figs. \ref{fig5}(a) and \ref{fig5}(b)]. Moreover, both the top valence band and the bottom of conduction band around the Dirac point are mainly contributed by the 2\emph{p}$_z$ orbitals of C and N2 atoms, while the 2\emph{p}$_z$ orbital of N1 atoms also makes a contribution to the lowest conduction band. These 2\emph{p}$_z$ orbitals overlap, resulting in the formation of an extended $\pi$-electron conjugation system in the C$_4$N$_3$H monolayer, similar to that in graphene. Accordingly, we can take the conclusion that this $\pi$-electron conjugation system is the reason for the emergence of Dirac states in this organic monolayer.

\section{\label{sec:level1}Conclusion}

To summarize, we designed a novel 2D organic material with evenly distributed heart-shaped angstrom-scale pores, namely the C$_4$N$_3$H monolayer. In its unique structure, every C atom is bonded with three other atoms (C or N atoms) \emph{via} \emph{sp}$^2$ hybridization, giving rise to three $\sigma$-like orbitals placed in its basal plane and one $\pi$ orbital along \emph{Z} axis in the perpendicular direction. These $\sigma$ and $\pi$ bonds are responsible for the energy stability of the 2D C$_4$N$_3$H framework. Furthermore, we have confirmed its dynamic, thermal and mechanical stabilities respectively by its phonon dispersion curves, FPMD simulations and mechanical properties. Band structure and TDOS clearly reveal that this organic monolayer is a semimetal with anisotropic Dirac cones and very high Fermi velocity which is roughly one order of magnitude larger than that in 2D organic Dirac materials ever reported \cite{17,18}. Based on the PDOS and the band decomposed charge density of this monolayer, we demonstrate that the anisotropic Dirac cones originate from the extended $\pi$-electron conjugation system which is mainly contributed by the 2\emph{p}$_z$ orbitals of C and N atoms. These results indicate that we successfully designed a 2D organic Dirac material with high Fermi velocity which can be comparable to that in 2D inorganic ones. We hope that our findings will promote the experimental realization of this novel material and greatly push forward the study of the 2D organic Dirac materials.

\begin{acknowledgments}
We thank Feng Liu of University of Utah for helpful discussion. This work was financially supported by the State Key Program for Basic Research (Grant Nos. 2014CB921102 and 2017YFA0206304) and NSFC (Grant Nos. 51572122 and 11304096), China. We are grateful to the High Performance Computing Center of Nanjing University for doing the numerical calculations in this paper on its blade cluster system.
\end{acknowledgments}


\end{document}